\documentstyle[pre,multicol,aps]{revtex}

\begin{document}
\setlength{\textwidth}{180mm}
\setlength{\textheight}{240mm}
\setlength{\parskip}{2mm}

\input{epsf.tex}
\title{Approximate solutions and scaling transformations 
       for quadratic solitons}
\author{Andrey A. Sukhorukov}

\address{Australian Photonics Cooperative Research Centre,  Research
School of Physical Sciences and Engineering \\  Optical Sciences Centre,
Australian National University, 
Canberra, Australian Capital Territory 0200, Australia}

\maketitle

\begin{abstract}
We study quadratic solitons supported by 
two- and three-wave parametric interactions in $\chi^{(2)}$
nonlinear media. 
Both planar and two-dimensional cases are considered.
We obtain very accurate, {\em 'almost exact'}, explicit analytical solutions, 
{\em matching the actual bright soliton profiles}, 
with the help of a specially-developed approach, based on analysis of
the scaling properties. Additionally, we use these approximations to describe
the linear tails of solitary waves which are related to
the properties of the soliton bound states.
\end{abstract}


\vspace*{-1.0cm}

\begin{multicols}{2}
\narrowtext

\section{Introduction}

One of the rapidly expanding areas of research is the physics of solitons -- 
wave packets, or self-trapped beams, that propagate with their profiles 
remaining undistorted.
In particular, {\em parametric solitons}, composed of mutually trapped
fundamental and harmonic waves, attract interest of researchers due
to a wide range of possible applications. 
In optics, for example, such
solitons were observed in media with quadratic (or $\chi^{(2)}$)
nonlinearity, 
and their unique features can be utilized for the creation of all-optical
information processing devices \cite{rev_stegeman}.
In general, parametric solitons may form in different media which 
possess resonant quadratic nonlinearities, such as
plasma, organic superlattices
\cite{rev_altern}, Bose-Einstein condensate \cite{BEC}, etc.

Many papers have been devoted to a theoretical analysis of quadratic
solitons \cite{rev_chi2}. It was shown that bright solitons can be stable,
and hence are of most interest for practical applications, 
on the other hand parametric 
dark solitons often exhibit modulation instability. However, due to 
nontrivial features of the resonant coupling, there still remain some 
properties of the bright quadratic solitons which have not been thoroughly 
described or completely understood.
The problem here is that the governing equations are {\em not integrable}, and
general analytical solutions can't be constructed. Thus, the
variational method \cite{var_m} was widely used to find approximate solutions.
In this approach, the free parameters controlling the trial functions are
found by minimizing the Lagrangian functional.
A rarely mentioned limitation here is that this technique imposes
very strong restrictions on the class of the trial functions, for the
parameters to be found in an explicit form. This becomes a real drawback if
these test functions are not quite suitable for the problem at hand 
(see, e.g., Ref \cite{aa}).

In this paper we introduce a {\em different approach for obtaining approximate
expressions for the soliton envelopes}. At first we choose trial functions 
which can precisely describe the actual soliton profiles. This step
involves the analysis of the scaling properties of a soliton family, i.e. how
the envelopes are transformed as a free parameter (propagation constant) is
altered.
Secondly, a specially developed technique is used to find the fitting
parameters. It will be demonstrated that the resulting solutions turn out
to be {\em extremely accurate}.

The rest of the paper is organized as following. First, two-wave solitons
in planar structures are studied. Then, the analysis is extended to
the case of three wave mutual trapping in an anisotropic medium. Finally,
the properties of two-component solitary beams propagating in a bulk
medium are investigated.

\section{One dimensional solitons}
\subsection{Two-wave solitons}
\subsubsection{Basic equations and their properties}

Parametric interaction between the fundamental frequency (FF) wave and its 
second harmonic (SH) in the (1+1) dimensional case 
can be described by a set of coupled equations for slowly varying
complex amplitudes of the wave packets 
\cite{f_ks} (see also \cite{rev_chi2,f_ksf}). 
We consider the case when there is
no walk-off, and then in normalized variables have \cite{sfam_bk}:
\begin{equation} \label{eq:uw_ns}
 \begin{array}{l}
  {\displaystyle 
   i \frac{\partial u}{\partial z} + 
  \frac{\partial^2 u}{\partial x^2} + u^* w  = 0,} 
          \\*[9pt]
  {\displaystyle 
   i \sigma \frac{\partial w}{\partial z} + 
  \frac{\partial^2 w}{\partial x^{2}} - \beta w + \frac{1}{2} u^2 = 0,}
  \end{array}
\end{equation}
where $u(x,z)$ and $w(x,z)$ are the FF and SH amplitudes correspondingly, 
$z$ is the propagation distance,
and the parameter $\beta$ characterizes the mismatch of the linear phase
velocities. 
These equations can describe: (i)~{\em spatial beams} in a slab waveguide, 
exhibiting diffraction in the transverse direction $x$, 
where $\sigma \approx 2$; 
or (ii)~{\em temporal pulses}, where $x$ stands for the
retarded time and $\sigma > 0$ is the ratio of the absolute values of 
second-order dispersion coefficients (the signs are those which allow
stable bright solutions).

Under certain conditions, mutual trapping of two waves can be achieved,
when diffraction (or second-order dispersion) is exactly compensated for 
by nonlinear refraction.
Such stationary propagation is observed for a special class of 
beams (or temporal wave packets)
-- {\em solitons}.
To find their profiles, we look for solutions of Eqs.~(\ref{eq:uw_ns}) in
the form
\begin{equation} \label{eq:uw0}
 \begin{array}{l}
  {\displaystyle 
    u(x,z) = \lambda u_0 \left( x \sqrt{\lambda} \right) e^{i \lambda z},} 
          \\*[9pt]
  {\displaystyle
    w(x,z) = \lambda w_0 \left( x \sqrt{\lambda} \right) e^{2 i \lambda z},}
  \end{array}
\end{equation}
where $u_0$ and $w_0$ are the real envelope amplitudes, and
$\lambda > 0$ is the propagation constant. After substituting 
expressions~(\ref{eq:uw0}) into Eqs.~(\ref{eq:uw_ns}), the
following system of coupled ordinary differential equations 
can be derived:
\begin{equation} \label{eq:uw}
 \begin{array}{l}
  {\displaystyle 
  \frac{d^2 u_0}{d x^2} - u_0 + u_0 w_0  = 0,} 
          \\*[9pt]
  {\displaystyle
  \frac{d^2 w_0}{d x^{2}} - \alpha w_0 + \frac{1}{2} u_0^2 = 0,}
  \end{array}
\end{equation}
where the only free parameter is the normalized mismatch 
$\alpha = 2 \sigma + \beta / \lambda$.

For localized waves, described by Eqs.~(\ref{eq:uw_ns}), the total power $P$ 
and Hamiltonian $H$ are conserved \cite{f_ks}. 
The values of these integral characteristics, corresponding to solitons 
defined by
Eqs.~(\ref{eq:uw}), can be found as \cite{krub}:
\begin{equation} \label{eq:HP}
  P = \lambda^{3/2} P_0, \,
  H = \lambda^{5/2} ( H_0 - P_0 ) .
\end{equation}
Here the renormalized power and Hamiltonian are 
$P_0 = P_{u_0} + 2 \sigma P_{w_0}$ and
$H_0 = 0.4 ( P_{u_0} + \alpha P_{w_0})$, where
\begin{equation} \label{eq:power1d}
 P_{u_0} = \int_{-\infty}^{+\infty} u_0^2 dx,\,
 P_{w_0} = \int_{-\infty}^{+\infty} w_0^2 dx, 
\end{equation}

We are looking for {\em bright solitons}, where the field decays at infinity. 
Such solutions of Eqs.~(\ref{eq:uw})
were found numerically for any $\alpha > 0$ 
\cite{sfam_bk,sfam_torner,sfam_drum} and 
corresponding solitons were shown to be stable for
mismatches $\alpha > \alpha_{\rm cr}$ \cite{stability}. Here, the critical
parameter value $\alpha_{\rm cr}$ is 
a function of $\sigma$, and
for spatial solitons 
${\left. \alpha_{\rm cr} \right|}_{ \sigma = 2 } \approx 0.2$.

The properties of this soliton family were extensively studied in the
literature.
An exact solution was found 
at $\alpha=1$ \cite{f_ks}:
\begin{equation} \label{eq:slv_al_1}
  w_0 (x) = u_0 (x) / \sqrt{2} = 
       {\left( 3/2 \right)} \; {\rm sech}^{2} \left( x/2 \right) .
\end{equation}
On the other hand, for large mismatches,
$\alpha \rightarrow \infty$, the SH component approaches 
$w_0(x) \simeq v_0^2 (x) / {\left( 2 \alpha \right)}$.
Then, in this so-called cascading limit, the FF wave is
determined as a solution of the nonlinear Schr\"odinger equation (NLSE), 
and the wave envelopes are \cite{f_ksf,nls_drum}:
\begin{equation} \label{eq:slv_al_inf}
  v_0(x) \simeq 2 \sqrt{\alpha} \; {\rm sech} (x),\; 
  w_0(x) \simeq 2 \; {\rm sech}^2 (x).
\end{equation}
This solution can be improved by taking into account higher-order terms in
a series decomposition over a small parameter, $\alpha^{-1}$ 
\cite{nls_berge,var_exp}.
However, this approach leads to cumbersome expressions which are somewhat
hard to analyze.

Exact analytical solutions of Eqs.~(\ref{eq:uw}) cannot be found for 
arbitrary values of $\alpha$. Thus, in order to obtain approximate solutions 
for the soliton profiles, the variational approach was
used. Calculations with an ansatz in the form of Gaussian functions 
predict the power distribution between the FF and SH components quite 
accurately, and provide a close estimation for the maximum amplitudes in the
whole parameter range $0 < \alpha < \infty$ \cite{var_exp,var_exp_bs}.
However, the trial functions do not correspond to 
the actual wave profiles, and thus the "tails", or amplitude asymptotics 
at $x \rightarrow \pm \infty$, are not described well.
In other studies \cite{var_sech},
the profiles of the trial functions are chosen as scaled exact 
solutions (\ref{eq:slv_al_1}) or (\ref{eq:slv_al_inf}) with arbitrary 
amplitudes, but fixed relative widths for the FF and SH wave packets. 
Due to this limitation, precise results 
are obtained only for $\alpha \approx 1$ and 
$\alpha \rightarrow \infty$ respectively.

\subsubsection{Approximate analytical solution}

In order to construct a solution without the above mentioned
drawbacks, we 
want to take into account 
some characteristic
properties of the bright solitons. 
Specifically, we notice that the form of the SH envelope
is the same at $\alpha = 1$ and $\alpha \rightarrow \infty$ 
[see Eqs.~(\ref{eq:slv_al_1}) and (\ref{eq:slv_al_inf})]. 
Moreover, we perform numerical simulations and
observe a very remarkable fact: 
for $\alpha \geq 1$, $w_0(x)$ remains almost
self similar. Thus, we search for an approximate solution with the SH
component in the form:
\begin{equation} \label{eq:SH_approx}
    w_0 (x) = w_m {\rm sech}^2 {\left( x / p \right)},
\end{equation}
where the maximum amplitude $w_m$ and characteristic width $p$ are 
{\em unknown parameters}. Then, the FF component profile can be determined 
using the first equation in (\ref{eq:uw}). This is a linear eigenvalue 
problem, which has an exact solution for the effective waveguide created by 
the SH field (\ref{eq:SH_approx}).
In a single bright soliton, $u_0(x)$ doesn't have zeros, 
and thus we take the fundamental mode. 
This requirement leads to a relation between
the parameters of an effective waveguide:
\begin{equation} \label{eq:p-wm}
  w_m = 1 + 1 / p.
\end{equation}
Then the corresponding FF profile is found to be
\begin{equation} \label{eq:FF_approx}
  u_0 (x) = u_m {\rm sech}^p {\left( x / p \right)},
\end{equation}
where $u_m$ is the peak amplitude. 

Our trial functions do not satisfy the SH equation in (\ref{eq:uw}) exactly, 
and thus it should be matched approximately.
This can be done with the help of 
the variational method.
However, this leads to a set of
transcendental equations, and then the solution parameters
can't be expressed in an explicit form. 
As our aim is to derive a simple analytical approximation, which should 
be easy to analyze and use in calculations, we choose another approach.
First, in order to match the soliton peak,
we require the equation for the SH amplitude in
(\ref{eq:uw}) to be exactly satisfied at the soliton center, $x=0$, and
thus obtain:
\begin{equation} \label{eq:SH_eq0}
  2 w_m {\left( \alpha + 2 / p^2 \right)} = u_m^2 .
\end{equation}
Secondly, we notice that Eqs.~(\ref{eq:uw}) describe an equivalent dynamical
problem, viz. particle movement with generalized velocities
$\left( d u_0 / d x, d w_0 / d x \right)$ in a potential 
$U_d(u_0,w_0) = {\left( u_0^2 w_0 - u_0^2 - \alpha w_0^2 \right)} / 2$. 
This is a conservative system, with Hamiltonian 
$H_d = {{\left( d u_0 / d x \right)}^2}/2 
  + {{\left( d w_0 / d x \right)}^2}/2 + U_d(u_0,w_0)$. 
For bright solitons, the field vanishes at infinity, and thus $H_d \equiv 0$. 
Then, as the functions $u_0(x)$ and $w_0(x)$ reach their
maximum values at $x=0$, the corresponding derivatives are zero, and thus
the necessary condition for a zero asymptotic is
$\left. H_d \right|_{x=0} = \left. U_d \right|_{x=0} = 0$, 
which we use to relate the peak amplitudes:
\begin{equation} \label{eq:U_eq0}
  u_m^2 w_m - u_m^2 - \alpha w_m^2 =0 .
\end{equation}
Combining Eqs.~(\ref{eq:SH_approx})-(\ref{eq:U_eq0}), we obtain an
approximate solution in the following simple form:
\begin{equation} \label{eq:slv2}
 \begin{array}{l}
  {\displaystyle 
    u_0(x) = {u_m}{{\rm sech}^{p}(x/p)}, \;\;\; 
    w_0(x) = {w_m}{{\rm sech}^{2}(x/p)},
   }  \\*[9pt]
  {\displaystyle 
    u_m^2 = \frac{\alpha w_m^2}{\left( w_m -1 \right)}, \;
    p= \frac{1}{\left( w_m-1 \right)}, \;
    \alpha = \frac{4 {\left( w_m-1 \right)}^3}{\left( 2-w_m \right)}.
  } \end{array}
\end{equation}
Here, the last relation allows us to determine $w_m$ for an 
arbitrary $\alpha$ as a solution of a cubic equation, 
and then to find all other parameters as functions of $\alpha$. 
For mismatches in the interval $0 < \alpha < +\infty$, the 
parameter values change monotonically in the ranges:
$0 < u_m < +\infty$, $1 < w_m < 2$, and $+\infty > p > 1$. At $\alpha=1$ the
values are $u_m = 3/\sqrt{2}$, $w_m = 3/2$, $p = 2$, i.e. our general 
expression reduces to the exact solution (\ref{eq:slv_al_1}). Similarly, for
$\alpha \rightarrow +\infty$ the limiting result (\ref{eq:slv_al_inf})
follows.

\subsubsection{Soliton tails}

\begin{figure}
 \setlength{\epsfxsize}{8cm}
 \centerline{\mbox{\epsffile{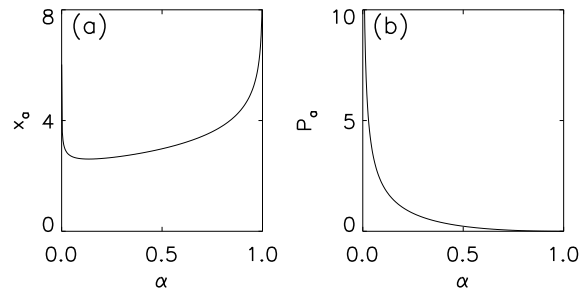}}}
 \vspace*{5mm}
 \caption{ \label{fig:ltail}
 Dependence of the SH linear tails characteristics on mismatch $\alpha$:
 (a)~separation from the soliton center $x_a$ and
 (b)~associated normalized power $P_a$.
 }
\end{figure}

As one of our goals is to describe the soliton tails, let us have a look 
at the far-field asymptotics that follow from Eqs.~(\ref{eq:slv2}). 
It is easy to check that the FF profile exactly matches
the linear limit.
To understand the properties of the SH component tails, 
we notice that the corresponding equation in Eqs.~(\ref{eq:uw}) 
describes the motion of a particle driven by an external force 
$u_0^2 /2 $. As this expression is positive, the function $w_0(x)$ can't
decay faster than that in the linear limit: 
$w_0(x) \sim \exp{\left( -\sqrt{\alpha} |x| \right)}$. Indeed, for 
$\alpha>1$, we have $p < 2$, i.e. the FF component is effectively wider 
than the SH, and thus the field decay rate is smaller, as is correctly 
predicted by Eqs.~(\ref{eq:slv2}). 
In contrast, for $\alpha < 1$ the width of the FF 
component is smaller than that of the SH (as $p < 2$), and then
the SH tails are to be determined by linear asymptotics.
However, the solution (\ref{eq:slv2}) 
overestimates the SH field localization. 
To account for this feature, we sue the soliton `peak' from
(\ref{eq:slv2}) with a linear tail, so that the function and its first
derivative change continuously,
and obtain a more accurate expression for the SH component 
in the case $\alpha < 1$:
\begin{equation} \label{eq:slv2ds}
 w_0(x) = \left\{ \begin{array}{l}
    {\displaystyle 
       w_m {\rm sech}^2 \left( x / p \right), 
                      \, {\left| x \right|} \le x_a , 
    } \\*[9pt]
    {\displaystyle 
       w_0(x_a) \exp{\left[ -\sqrt{\alpha} 
          {\left( {\left| x \right|} - x_a\right)} \right]}, 
                      \,  {\left| x \right|} > x_a ,
   } \end{array} \right.
\end{equation}
where $x_a = {\left( p / 2 \right)} 
            \ln{\left[ \left(2 + p \sqrt{\alpha}\right) / 
                        \left(2 - p \sqrt{\alpha}\right)\right]}$. 
From Eq.~(\ref{eq:slv2ds}) it follows 
that the SH profile becomes {\em double scaled}.
That is, we predict the existence of {\em linear tails} in the SH component, 
which are effectively not trapped by the FF field, and carry some power:
\begin{equation} \label{eq:lt_power}
   P_a = 2 \int_{x_a}^{+\infty} w_0^2 dx = 
                      [w_0(x_a)]^2 / \sqrt{\alpha}.
\end{equation}
The dependence of the linear tail parameters on the normalized
mismatch $\alpha$ is shown in Fig.~\ref{fig:ltail}.

\subsubsection{Soliton bound states}

A connection between the soliton bound states and the linear tails has been 
demonstrated for many physical situations.
For example, mutual trapping of radiating parametric bright solitons 
was shown to be possible for a discrete set of propagation constant values 
when, due to destructive interference, the oscillatory tails 
disappear \cite{bs_bur}. 
However, in our case the localized modes are not in resonance with 
propagating linear waves \cite{sfam_bk}, 
and a natural assumption is that the
{\em solitons can trap each other by their linear tails} for a continuous
range of mismatches. This physical picture is consistent with 
the results of the previous numerical simulations 
and analytical investigations,
showing that multi-soliton states are possible only for $\alpha < 1$ 
\cite{sfam_drum,var_exp_bs,bs_ml,bs_champ}.
Moreover, the characteristic distance between the neighboring solitons 
can be roughly estimated to be of order $2 x_a$, and this expression 
predicts the {\em non-monotonic dependence 
of the separation on mismatch $\alpha$}. 
The minimum separation should be observed for the mismatch 
corresponding to an extremum point, 
$\left. d x_a / d \alpha \right|_{\alpha_{\rm ma}} = 0$, 
and then it follows that
$\alpha_{\rm ma} \simeq 0.12$, see Fig.~\ref{fig:ltail}(a). 
Quite remarkably, this mismatch value
corresponds very closely to the results of numerical calculations,
see Fig.~4 in Ref.~\cite{bs_champ}.

\begin{figure}
 \setlength{\epsfxsize}{8cm}
 \centerline{\mbox{\epsffile{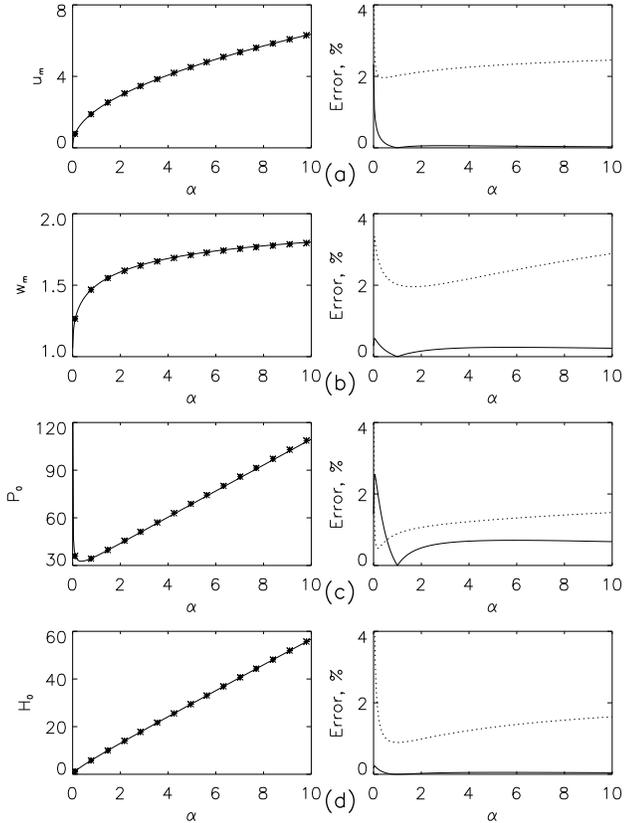}}}
 \vspace*{5mm}
 \caption{ \label{fig:compar_1d2w}
 Left: Comparison between numerical results (continuous curves) and
 approximate analytical solutions (crosses) given 
 by Eqs.~(\ref{eq:slv2})-(\ref{eq:slv2ds}).
 The characteristics are: 
 (a,b)~maximum amplitudes of the FF and SH, 
 (c)~total power (for the case $\sigma = 2$), and 
 (d)~Hamiltonian. \\
 Right: Corresponding relative errors are shown with solid lines, and dotted
 lines present deviations for the SGP.
 }
\end{figure}

\subsubsection{Comparison with numerical results}

In order to determine deviations between the approximate and exact
solutions, we solved Eqs.~(\ref{eq:uw}) numerically. 
The corresponding dependencies of the
peak amplitudes, total power, and Hamiltonian 
on the detuning parameter $\alpha$ are
presented in Fig.~\ref{fig:compar_1d2w} (left graphs) 
for the case of spatial solitons ($\sigma = 2$). 
As a matter of fact, the numerical and analytical 
results on these plots are {\em not distinguishable},
and that is why we show them differently, by continuous curves and asterisks. 
Plots on the right give corresponding relative deviations with
solid lines, and dashed lines demonstrate the errors for the 
variational solution with Gaussian profiles (SGP) \cite{var_exp,var_exp_bs}.
We see that the analytical solution (\ref{eq:slv2})-(\ref{eq:slv2ds}) 
{\em describes the key soliton parameters extremely accurately}. 
In a wide region of mismatch values, $\alpha > 1$,
the relative errors are smaller than 0.7\% for the total power, 
and 0.3\% for other characteristics.
For $\alpha < 1$ the deviations become larger, but do not exceed 3\%. 
It is clear that our solution gives much more accurate results than the SGP,
and the latter provides a slightly better estimation for the total power
only in a narrow region of mismatches.

\begin{figure}
 \setlength{\epsfxsize}{8.0cm}
 \centerline{\mbox{\epsffile{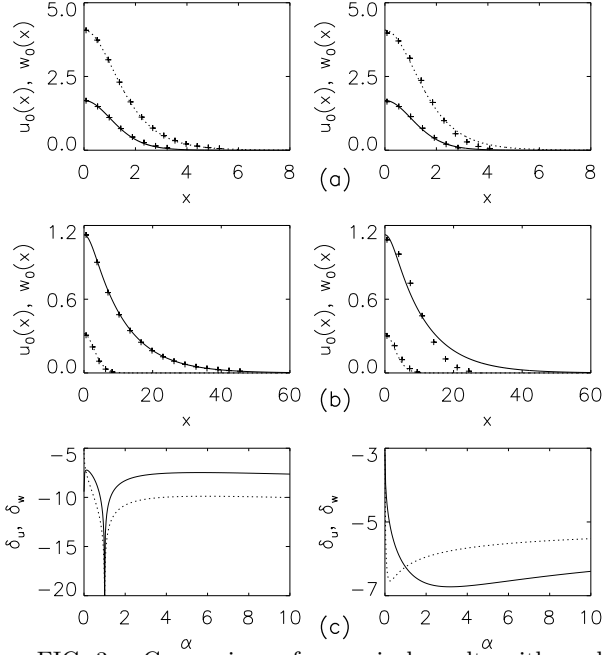}}}
 \vspace{2mm}
 \caption{ \label{fig:prof1d2w}
 Comparison of numerical results with analytical predictions:
  solution (\ref{eq:slv2})-(\ref{eq:slv2ds}), on the
 left, and the SGP, on the right. Dotted lines correspond to the FF and 
 solid lines to the SH components.
 (a,b)~Two-wave soliton profiles at \mbox{$\alpha=4$} and 
       \mbox{$\alpha = 0.01$} respectively.
 Continuous curves show exact numerical and crosses show
 approximate solutions.
 (c)~Deviations of approximate profiles as defined in Eqs.~(\ref{eq:uw_err}).
}
\end{figure}

At this point we would like to stress that the 
above-mentioned characteristics are not the only ones which 
determine the `quality' of the approximate solutions. 
The closeness of the approximate component profiles to exact soliton
envelopes,
including the tails, 
is also very important.
From Figs.~\ref{fig:prof1d2w}(a) and \ref{fig:prof1d2w}(b), 
we see that the solution (\ref{eq:slv2})-(\ref{eq:slv2ds})
describes the profiles very accurately as well (left plots), 
unlike the SGP which matches them only `on average' 
(see plots on the right).
To characterize the accuracy numerically, 
we define the relative deviations as:
\begin{equation} \label{eq:uw_err}
 \begin{array}{l}
  {\displaystyle 
   \delta_u = \ln { 
       \frac{\int_{-\infty}^{+\infty} 
             {\left| u_0^{\rm exact} - u_0^{\rm approx.} \right|}^2 dx}
            {\int_{-\infty}^{+\infty} 
             {\left| u_0^{\rm exact} \right|}^2 dx} 
       }, }
          \\*[15pt]
  {\displaystyle
   \delta_w = \ln {
         \frac{\int_{-\infty}^{+\infty} 
               {\left| w_0^{\rm exact} - w_0^{\rm approx.} \right|}^2 dx} 
              {\int_{-\infty}^{+\infty} 
               {\left| w_0^{\rm exact} \right|}^2 dx} 
       }.}
  \end{array}
\end{equation}
Note that these characteristics have a dB-like scale, i.e. smaller negative
values mean better matching.
In Fig.~\ref{fig:prof1d2w}(c), the errors corresponding to solution 
(\ref{eq:slv2})-(\ref{eq:slv2ds}) are shown on the left, and to SGP on the
right. We see that our approach allows us to {\em precisely describe the 
soliton profiles, both peaks and tails, for any mismatch $\alpha$}.
That is why we were able to reveal some remarkable features of two-component 
parametric mutual trapping for $\alpha < 1$, when the SH field configuration 
becomes double scaled [see Eq.~(\ref{eq:slv2ds}),(\ref{eq:lt_power}) and 
relating discussions].
On the other hand, the SGP does not provide close estimations for the actual 
profiles, especially for $\alpha < 1$, where the discrepancies increase 
drastically.

\begin{figure}
 \setlength{\epsfxsize}{8.0cm}
 \centerline{\mbox{\epsffile{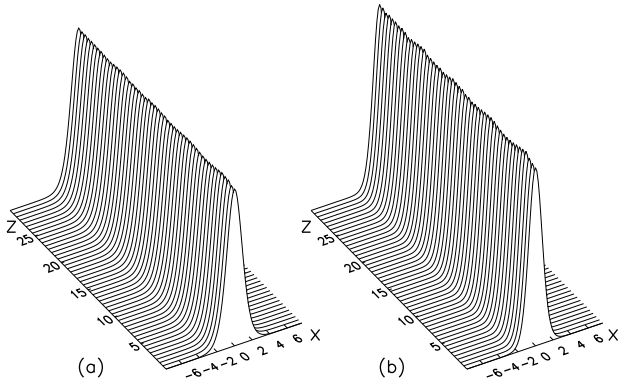}}}
 \vspace{2mm}
 \caption{ \label{fig:uw_prop}
 Almost stationary evolution of the FF (a) and SH (b). The initial condition
 is given by the approximate two-wave soliton solution (\ref{eq:slv2}) at
 $\alpha = 4$, $\sigma = 2$, $\lambda=1$. 
 }
\end{figure}

Approximate solutions can be used not only in 
theoretical studies, but also in numerical simulations.
The unique accuracy of solution (\ref{eq:slv2})-(\ref{eq:slv2ds}) makes it an 
almost {\em perfect generator for `soliton' input conditions}. 
We checked that the initial propagation stage is accompanied by very minor 
oscillations, and that the associated power loss due to radiation 
is negligible [see, e.g., Fig.~\ref{fig:uw_prop}]. 

To summarize, the approximate analytical solution, given in a compact
explicit form by Eqs.~(\ref{eq:slv2})-(\ref{eq:slv2ds}), 
describes virtually all principal features of the soliton family.
That is why 
it can be called {\em `almost exact'}. 
We might wonder, why is it so accurate? 
The key point in the analysis was to take into account the self-similarity 
of the SH envelopes.
Then,
we can view Eqs.~(\ref{eq:slv2})-(\ref{eq:slv2ds}) 
as {\em approximate scaling transformations} of the two-wave bright 
soliton family.

\subsection{Three-wave interaction in anisotropic medium}

Let us now investigate a more general case of three-wave interaction.
Following Ref.~\cite{us}, we consider a double-phase matched wave 
interaction, such that the FF waves of fundamental orthogonal polarizations 
(i.e. ordinary and extraordinary)
are coupled with the same SH component. 
That is, the full FF field is now vectorial, 
$\overrightarrow{U} = \{ u,v \}$, and
the two components are determined as 
\begin{equation} \label{eq:polarize}
 \begin{array}{l}
  {\displaystyle 
    u(x,z) = U(x,z) \cos \left[ \varphi(x,z) \right] ,} 
                     \\*[9pt]
  {\displaystyle 
    v(x,z) = U(x,z) \sin \left[ \varphi(x,z) \right] ,}
  \end{array}
\end{equation}
where $\varphi(x,z)$ is the polarization angle. For such a configuration 
the system of normalized equations can be written in the following 
form \cite{us}:
\begin{equation} \label{eq:uvw_ns}
 \begin{array}{l}
  {\displaystyle 
   i \frac{\partial u}{\partial z} + 
    \frac{\partial^{2}u}{\partial x^{2}}  + u^{\ast} w = 0,} 
                     \\*[9pt]
  {\displaystyle 
   i \frac{\partial v}{\partial z} + 
    \frac{\partial^{2} v}{\partial x^{2}} - \beta_1 v + \chi v^{\ast}w= 0,} 
                     \\*[9pt]
  {\displaystyle 
   i \sigma \frac{\partial w}{\partial z} + 
    \frac{\partial^{2}w}{\partial x^{2}} - \beta w+\frac{1}{2}(u^2+v^2)= 0.}
  \end{array}
\end{equation}
Here $\chi$ is the normalized component of the $\chi^{(2)}$ nonlinear 
susceptibility matrix, characterizing the $v \leftrightarrow w$ 
coupling `strength' in relation to the $u \leftrightarrow w$ parametric 
interaction process, and $\beta_1$ is the phase mismatch between
the orthogonally-polarized FF components. All the other parameters have the
same meaning as in Eqs.~(\ref{eq:uw_ns}).
 
We are interested in bright solitons supported by  
Eqs.~(\ref{eq:uvw_ns}), and search for solutions in the form:
\begin{equation} \label{eq:uvw0}
 \begin{array}{l}
  {\displaystyle 
    u(x,z) = \lambda u_0( x \sqrt{\lambda} ) e^{i \lambda z},} 
                     \\*[9pt]
  {\displaystyle 
    v(x,z) = \lambda v_0( x \sqrt{\lambda} ) e^{i \lambda z},} 
                     \\*[9pt]
  {\displaystyle 
    w(x,z) = \lambda w_0( x \sqrt{\lambda} ) e^{2 i \lambda z},}
  \end{array}
\end{equation}
where $u_0$ and $v_0$ are the real stationary amplitudes of the FF waves with 
orthogonal polarizations, $w_0$ is the envelope of the SH component, 
and $\lambda > 0$ is the propagation constant. 
Then, we substitute Eqs.~(\ref{eq:uvw0}) into the original system 
(\ref{eq:uvw_ns}) and find a set of coupled equations for the soliton
profiles:
\begin{equation} \label{eq:uvw}
 \begin{array}{l}
  {\displaystyle 
  \frac{d^2 u_0}{d x^2} - u_0 + u_0 w_0  = 0,} 
          \\*[9pt]
  {\displaystyle 
  \frac{d^2 v_0}{d x^2} - \alpha_1 v_0 + \chi v_0 w_0  = 0,} 
          \\*[9pt]
  {\displaystyle 
  \frac{d^2 w_0}{d x^{2}} - \alpha w_0 + \frac{1}{2} (u_0^2 +v_0^2) = 0,}
  \end{array}
\end{equation}
where the normalized mismatches are
\begin{equation} \label{eq:al1_lambda}
 \alpha = 2 \sigma + \beta / \lambda,\,\,\,
 \alpha_1 = 1 + \beta_1 / \lambda.
\end{equation}

Similarly to the two-wave case, the total power $P$ and Hamiltonian $H$ 
of system (\ref{eq:uvw_ns}) can be found for stationary solutions using
Eqs.~(\ref{eq:HP},\ref{eq:power1d}), where
$P_0 = P_{u_0} + 2 \sigma P_{w_0} + (1/\chi) P_{v_0}$,
$H_0 = 0.4 [ P_{u_0} + \alpha P_{w_0} + (\alpha_1/\chi) P_{v_0} ]$, and
$P_{v_0} = \int_{-\infty}^{+\infty} v_0^2 dx$.

First, let us study some general properties of Eqs.~(\ref{eq:uvw}). 
We notice that in an isotropic medium, $\chi = \alpha_1 = 1$ and 
the problem reduces to the two-wave case considered above, as the
FF wave can have an arbitrary constant polarization angle $\varphi$. 
Simple two-wave solutions are also possible in an anisotropic medium, but only 
for trivial polarizations: $\varphi = 0$ ($v_0 \equiv 0$) and 
$\varphi = \pi / 2$ ($u_0 \equiv 0$). However, it was shown that solitons 
with mixed polarizations also exist \cite{us}. To study such 
three-wave parametric coupling, we follow the same path as in the previous
section, and, in order to understand the principal scaling properties,  
refer to an exact one-parameter family of solutions found
for $\alpha_1=1/4$, $\chi=1/3$, and $\alpha > 1$ \cite{us}:
\begin{equation} \label{eq:uvw_slv_fam}
 \begin{array}{l}
  {\displaystyle 
   u_0(x) = {\left( 3/\sqrt{2} \right)}\, {\rm sech}^2\left( x/2 \right),} 
          \\*[9pt]
  {\displaystyle 
   v_0(x) = \sqrt{3 \left( \alpha-1 \right)}\, {\rm sech}\left( x/2 \right),} 
          \\*[9pt]
  {\displaystyle 
   w_0(x) = \left( 3/2 \right)\, {\rm sech}^2\left( x/2 \right).}
  \end{array}
\end{equation}
An interesting feature of this three-component solution is
that the SH profile is the same 
for any $\alpha$.
Thus, we suppose that the SH envelope shape always remains close 
to ${\rm sech}^2 (x/p)$, just as for two-component solitons,
and choose the trial function as in Eq.~(\ref{eq:SH_approx}).
The SH acts as an effective waveguide simultaneously for two different 
FF waves, and that is why, when solving the corresponding equations 
in Eqs.~(\ref{eq:uvw}), we obtain two relations between the SH profile 
characteristics:
\begin{equation} \label{eq:pq-wm}
  w_m = 1 + p^{-1} = \frac{ q (q+1) }{ \chi p^2 },
\end{equation}
where $q = \sqrt{\alpha_1} p$. 
From Eq.~(\ref{eq:pq-wm}), it is straightforward to find the parameters:
\begin{equation} \label{eq:slv3_pqw}
  w_m = \frac{ \alpha_1 - \sqrt{\alpha_1} }{ \chi - \sqrt{\alpha_1} },
  \quad
  p = q \alpha_1^{-1/2} = \frac{\chi - \sqrt{\alpha_1}}{\alpha_1 - \chi}.
\end{equation}
Then, the FF envelopes corresponding to these values are:
\begin{equation} \label{eq:uv_approx}
  u_0 (x) = u_m {\rm sech}^p {\left( x / p \right)},\;
  v_0 (x) = v_m {\rm sech}^q {\left( x / p \right)},
\end{equation}
where $u_m$ and $v_m$ are the peak amplitudes.
We would like to note
that in the frames of our approach, the SH
profile does not depend on $\alpha$, as follows from Eqs.~(\ref{eq:slv3_pqw}).
This is quite an interesting approximate
scaling property of the three-wave solitons, and it is exactly satisfied
for the solution presented in Eqs.~(\ref{eq:uvw_slv_fam}).

Now we have to determine the remaining unknown parameters, viz. the FF 
amplitudes $u_m$ and $v_m$. 
As for the two-wave case, we fulfill the SH equation at $x=0$:
\begin{equation} \label{eq:uvw_SH0}
  2 w_m {\left( \alpha + 2\; p^{-2} \right)} = u_m^2 + v_m^2.
\end{equation}
Then, we consider an equivalent Hamiltonian dynamic problem and, after
matching the values ${\left. H_d \right|}_{x=0} = 
{\left. H_d \right|}_{x \rightarrow +\infty} = 0$, obtain another relation:
\begin{equation} \label{eq:uvw_H0}
  u_m^2 {\left( w_m - 1 \right)} 
    + v_m^2 {\left( w_m - \alpha_1 \chi^{-1} \right)} - \alpha w_m^2 =0 .
\end{equation}

It is now convenient to turn back to the polar notations
(\ref{eq:polarize}). The total FF intensity and polarization angle at the 
soliton center can be found using Eqs.~(\ref{eq:uvw_SH0}),(\ref{eq:uvw_H0}):
\begin{equation} \label{eq:slv3_Uphi}
 \begin{array}{l}
  {\displaystyle 
    U_m^2 = 2 w_m \left( \alpha + 2\; p^{-2} \right),}
          \\*[9pt]
  {\displaystyle 
    \sin^2 \left. \varphi_m \right. = 
     {\left( 1 - \alpha_1 \chi^{-1} \right)}^{-1}
     {\left( \alpha w_m^2 U_m^{-2} - w_m + 1 \right)} .}
  \end{array}
\end{equation}

Finally,
approximate three-wave soliton profiles are given by 
Eqs.~(\ref{eq:SH_approx}),(\ref{eq:uv_approx}), with the parameters
in an explicit form from
Eqs.~(\ref{eq:slv3_pqw}),(\ref{eq:slv3_Uphi}).
With no lack of
generality, we hereafter assume that $\chi < 1$, as it is always possible to
swap the functions $u$ and $v$ before re-normalizing the physical equations
from which the system (\ref{eq:uvw_ns}) originates. Analysis reveals
that three-wave solutions exist
if the mismatches $\alpha$ and $\alpha_1$ fulfill the following inequality:
\begin{equation}  \label{eq:alpha_limits}
 \alpha^{(u)} (\alpha_1) < \alpha < 
  \left\{ \begin{array}{ll}
         \alpha^{(v)} (\alpha_1) , & q(\alpha_1) > 1, \\*[8pt]
         +\infty , & q(\alpha_1) \le 1, 
          \end{array} \right.
\end{equation}
where
\begin{equation} \label{eq:alpha_uv}
 \alpha^{(u)} = \frac{4 }{p^2 \left(p-1\right)},\,
 \alpha^{(v)} = \frac{4 \alpha_1}{q^2 \left(q-1\right)}.
\end{equation}
The corresponding polarization angles are
$\varphi \left[ \alpha \rightarrow \alpha^{(u)} \right] \rightarrow 0$ 
(i.e. $v \rightarrow 0$) and
$\varphi\left[ \alpha \rightarrow \alpha^{(v)} \right] \rightarrow \pi/2$ 
($u \rightarrow 0$).
It is now obvious that the {\em boundaries (\ref{eq:alpha_uv})
correspond to bifurcations} from a two-wave soliton 
to a three-wave one. 

\begin{figure}
 \setlength{\epsfxsize}{8cm}
 \centerline{\mbox{\epsffile{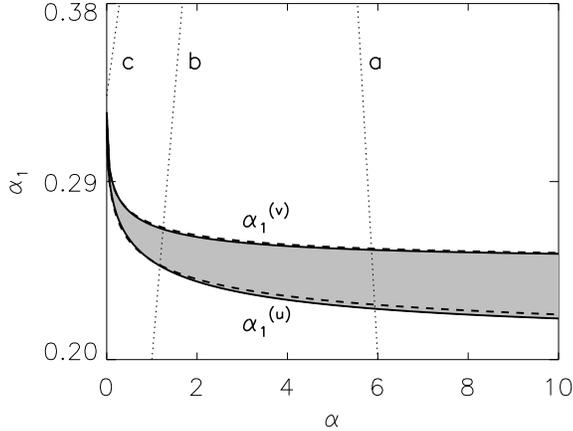}}}
 \vspace*{5mm}
 \caption{ \label{fig:bs_al_al1}
 Shaded region shows the numerically-found existence region for
 three-wave solitons; analytically calculated boundaries 
 (\ref{eq:alpha_uv}) are plotted with dashed lines.
 Dotted lines (a,b,c) show dependence of the normalized mismatches on the
 propagation constant $\lambda$ for $\beta_1=-4$ and $\beta=10, -15, -25$
 respectively. Parameters are $\chi = 1/3$ and $\sigma = 2$.
 }
\end{figure}

An example of the parameter region $(\alpha,\alpha_1)$,
where three-wave mutual trapping occurs,
is shown in Fig.~\ref{fig:bs_al_al1} for $\chi = 1/3$. 
We see that analytically calculated boundaries 
given by Eqs.~(\ref{eq:alpha_uv}) (shown with dashed lines) are 
extremely accurate, and almost
coincide with those found using numerical simulations.
Note that it also gives the correct prediction that
at $\alpha_1 \rightarrow \chi$ 
the two boundaries merge, i.e. $\alpha^{(u,v)} \rightarrow 0$.

\begin{figure}
 \setlength{\epsfxsize}{8cm}
 \centerline{\mbox{\epsffile{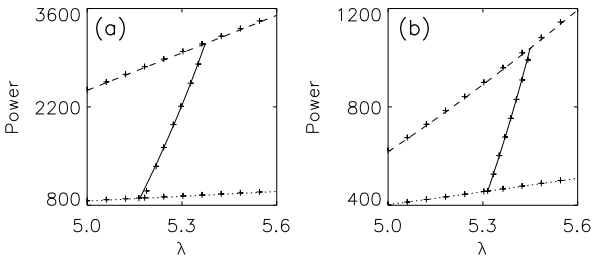}}}
 \vspace*{3mm}
 \caption{ \label{fig:pwr_3w}
 Dependence of the total soliton power $P$ on the propagation constant.
 Parameter values for the plots (a,b) are the same as in
 Fig.~\ref{fig:bs_al_al1} for the trajectories (a,b), respectively.
 In both cases, a three-wave soliton (solid line) 
 bifurcates from 
 two-wave solutions with $v=0$ (dotted) or $u=0$ (dashed). Crosses show
  approximate analytical results.
 }
\end{figure}

To understand the features of the three-wave solitons, we now recall that 
solutions of the original system (\ref{eq:uvw_ns}) constitute a one-parameter 
family, as follows from Eqs.~(\ref{eq:uvw0}). Then, according to 
Eq.~(\ref{eq:al1_lambda}), a change of the propagation constant $\lambda$ 
corresponds to motion along a
straight line in the parameter space $(\alpha,\alpha_1)$. The
limiting point for $\lambda \rightarrow +\infty$ is $(2 \sigma, 1)$, which
always lies above the three-wave existence region. 
Thus, we 
find that the trajectory will
go through this region only if $\beta_1 < 0$ and
$\beta > 2 \sigma \beta_1 / (1-\chi)$. Note that for fixed $\beta_1$ 
the value of $\beta$ determines the inclination angle, 
as demonstrated in Fig.~\ref{fig:bs_al_al1} by the lines (a,b). 
On the other hand, line (c) corresponds to the
case when the specified conditions do not hold, i.e. three-wave 
trapping is not possible for any $\lambda$.
From this analysis it follows that the three-wave soliton always 
bifurcates from a two-wave state with $v=0$, and then transforms 
into the other two-wave mode with $u=0$. This process can be easily 
seen in Fig.~\ref{fig:pwr_3w}, where examples of the power 
dependence on the propagation constant are shown. 

\vspace*{-2mm}

\begin{figure}
 \setlength{\epsfxsize}{8cm}
 \centerline{\mbox{\epsffile{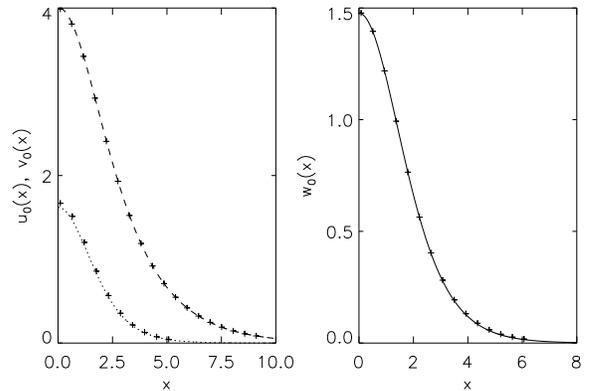}}}
 \vspace*{5mm}
 \caption{ \label{fig:prof_3w}
 Three-wave soliton profiles. 
 Left:~FF components ($u_0(x)$ -- dotted, and $v_0(x)$ -- dashed). 
 Right:~SH envelope.
 Crosses give analytically-calculated amplitudes.
 Parameters are the same as in Fig.~\ref{fig:pwr_3w}(a), and
 $\lambda = 5.35$.}
\end{figure}

The plots in Fig.~\ref{fig:pwr_3w} demonstrate 
that the analytical solution gives a very precise estimate for
the total power. It also accurately describes the soliton profiles in a wide
region of mismatches, with an example being shown in Fig.~\ref{fig:prof_3w}.
A thorough qualitative comparison is, however, a separate task,
as there exist several parameters which control the mutual trapping.
It will be presented elsewhere.

To summarize, we have analyzed
three-wave solitons in an anisotropic medium. Our approach allowed us to
predict corresponding parameter regions, and power dependence with high
accuracy.
Of course, some interesting aspects remain to be investigated,
such as stability, formation of linear tails, and properties of higher-order
modes. 
These problems are topics of separate study and thus will
not be addressed here, but the results obtained form a 
background for further in-depth investigations.

\section{Two-dimensional solitons}

Two-wave parametric spatial solitons can exist both in planar waveguides and
in bulk nonlinear media \cite{rev_chi2}.
In the latter case, the interaction between the FF and SH waves 
is described by the following system of coupled 
equations \cite{f_ks,var_exp,s2d_hk,s2d_fam}:
\begin{equation} \label{eq:uw2_ns}
 \begin{array}{l}
  {\displaystyle 
   i \frac{\partial u}{\partial z} + 
  \frac{\partial^2 u}{\partial x^2} + \frac{\partial^2 u}{\partial y^2} 
     + u^* w  = 0,} 
          \\*[9pt]
  {\displaystyle 
   i \sigma \frac{\partial w}{\partial z} + 
  \frac{\partial^2 w}{\partial x^{2}} + \frac{\partial^2 w}{\partial y^{2}} 
     - \beta w + \frac{1}{2} u^2 = 0,}
  \end{array}
\end{equation}
where $x$ and $y$ are the transverse coordinates. Other variables and
parameters are the same as for the (1+1) dimensional case described by 
Eqs.~(\ref{eq:uw_ns}), with the obvious difference that the amplitudes 
depend on three coordinates,
$u = u(x,y,z)$ and $w = w(x,y,z)$. We consider a spatial case
close to the phase matching, i.e. $\sigma \simeq 2$.

Our goal in this section is to calculate the soliton
envelopes by extending the method introduced for a planar case.
Specifically, we look for
circularly-symmetric stationary solutions of Eqs.~(\ref{eq:uw2_ns}), which
can be found with the help of the following substitution:
\begin{equation} \label{eq:uw20}
 \begin{array}{l}
  {\displaystyle 
    u(x,y,z) = \lambda u_0 \left( r \sqrt{\lambda} \right) e^{i \lambda z},} 
          \\*[9pt]
  {\displaystyle 
    w(x,y,z) = \lambda w_0 \left( r \sqrt{\lambda} \right) e^{2 i \lambda z},}
  \end{array}
\end{equation}
where $r = \sqrt{x^2 + y^2}$ is the radial distance in cylindrical coordinates, 
and $u_0(r)$ and $w_0(r)$ are the real normalized envelope functions.
Then, Eqs.~(\ref{eq:uw2_ns}) reduce to 
\begin{equation} \label{eq:uw2}
 \begin{array}{l}
  {\displaystyle 
  \frac{d^2 u_0}{d r^2} + \frac{1}{r} \frac{d u_0}{d r}
           - u_0 + u_0 w_0  = 0,} 
          \\*[9pt]
  {\displaystyle 
  \frac{d^2 w_0}{d r^2} + \frac{1}{r} \frac{d w_0}{d r}
          - \alpha w_0 + \frac{1}{2} u_0^2 = 0,}
  \end{array}
\end{equation}
where the propagation constant $\lambda$ and the normalized mismatch $\alpha$ 
are introduced in the same way as in Eqs.~(\ref{eq:uw0}),(\ref{eq:uw}).
Then, the soliton power and Hamiltonian are expressed via the
the normalized values as:
\begin{equation} \label{eq:HP2}
  P = \lambda P_0, \,
  H = \lambda^{2} ( H_0 - P_0 ) ,
\end{equation}
where $P_0 = P_{u_0} + 2 \sigma P_{w_0}$, 
and
\begin{equation}
 \begin{array}{rl}
   {\displaystyle
     P_{u_0} = 2 \pi \int_{0}^{+\infty} 
   } & {\displaystyle
        u_0^2 \; r dr,\,
      P_{w_0} = 2 \pi \int_{0}^{+\infty} w_0^2 \; r dr, } 
                            \\*[9pt]  
 {\displaystyle
  H_0 = 2 \pi \int_{0}^{+\infty} 
   } & {\displaystyle
      {\left[ 
          {\left( \frac{\partial u_0}{\partial r} \right)}^2
        + {\left( \frac{\partial w_0}{\partial r} \right)}^2  + \right.}  } 
                            \\*[9pt]  
     & {\displaystyle \, {\left.   
            + u_0^2
            + \alpha w_0^2
            - {u_0^2 w_0}
           \right]} \; r dr .
    } \end{array}
\end{equation}

Similarly to the one-dimensional case, Eqs.~(\ref{eq:uw2}) are not integrable,
and approximate soliton profiles were found 
using a variational method. 
Solutions were obtained for trial functions with
Gaussian profiles (SGP) and described dependence of the Hamiltonian and 
power on mismatch $\alpha > 0$ with a reasonable accuracy \cite{var_exp}.
The limitations of the SGP remain the same -- rough matching of actual
profiles and inadequate description of the
soliton tails. In order to construct a better solution, 
we have to start with some approximate
scaling property of the soliton family. The problem is that effective 
`dissipation' terms $\left( \sim \frac{1}{r} \frac{d}{d r} \right)$ 
in Eqs.~(\ref{eq:uw2}) lead to: 
(i)~distortion of soliton profiles in the center and 
(ii)~higher far field localization, which, taken together, 
result in very complicated scaling features. 
This makes a general analysis quite difficult,
and thus we limit our study to mismatches
$\alpha >1$. It has been shown that in this parameter range
the FF and SH relative beam width changes do not exceed 
a factor of two \cite{var_exp,s2d_fam}.
On the other hand, from the structure of Eqs.~(\ref{eq:uw2}), it follows that
the balance between the second-order and effective
dissipation linear terms, which affects the soliton shape, 
depends mainly on same characteristic beam width.
Thus, we can assume that, to some extent, the SH envelope form stays
almost intact for $\alpha > 1$, and the trial function is
\begin{equation} \label{eq:SH2_approx}
    w_0 (x) = w_s F\left( b_s r \right),
\end{equation}
where the function $F$ describes a characteristic SH profile, and
the relative amplitude $w_s$, together with the inverse width $b_s$, are 
scaling parameters. Then, we assume that, similarly to the (1+1)D
case (\ref{eq:FF_approx}), the FF profile can be approximately described as
\begin{equation} \label{eq:FF2_approx}
    u_0 (x) = u_s F^p {\left( b_s r \right)},
\end{equation}
where $p$ is an unknown parameter. 

To determine the SH characteristic
profile, we consider a mismatch $\alpha = 1$.
This case is easier to analyze, as
the envelopes of both mutually trapped 
components coincide, $w_0(r) = u_0(r) / \sqrt{2} = F(r)$. 
Here the function $F(r)$ satisfies the following equation
\begin{equation} \label{eq:Feq}
  \frac{d^2 F}{d r^2} + \frac{1}{r} \frac{d F}{ d r} 
       - F + F^2 = 0,
\end{equation}
whose approximate solution was found earlier with the help of a Hartree-like 
approach \cite{s2d_hk}. However, here we choose to use a variational
method in order to select the one which provides better matching.
First, we present Eq.~(\ref{eq:Feq}) in a variational form:
\begin{equation} \label{eq:Feq_var}
  \frac{ \delta L}{\delta F} = 0,
\end{equation}
where $\delta$ denotes the variational derivative, and $L$ is the Lagrangian
corresponding to the original Eq.~(\ref{eq:Feq}):
\begin{equation}
  L(F) = \int_0^{+\infty} {\left[
       {\left( \frac{d F}{d r} \right)}^2
       + F^2 - \frac{2}{3} F^3 
           \right]} r\;dr .
\end{equation}
Next, we select a trial function in the form 
\begin{equation} \label{eq:F0}
  F_0(r) = F_m {\rm sech}^2 ( b_0 r ),
\end{equation}
assuming that the profile is more or less close to that of planar solitons
(\ref{eq:SH_approx}). Now we note that, according to (\ref{eq:Feq_var}),
the Lagrangian reaches a minimum at an exact solution, and thus, to determine 
the peak amplitude $F_m$ and inverse width $b_0$ in the approximate expression 
(\ref{eq:F0}), an extremum point of the $L(F_0)$ integral should be found:
 $ {\partial L(F_0)}/{\partial F_m} = {\partial L(F_0)}/{\partial b_0} = 0$.
Solving these equations, we obtain the parameter values:
\begin{equation} \label{eq:F0_param}
 \begin{array}{l}
   {\displaystyle
     F_m = \frac{ 15 \left( 4 \ln 2 - 1 \right) }{ 32 \ln 2 - 11 }
         \approx 2.3781 ,
   } \\*[18pt] {\displaystyle
     b_0 = \frac{1}{2} \sqrt{ 
          \frac{ 5 \left( 4 \ln 2 - 1 \right) }{ 8 \ln 2 + 1 } }
         \approx 0.5818 .
   } \end{array}
\end{equation}
We made a comparison between the exact numerical and 
approximate solutions of Eq.~(\ref{eq:Feq}) given by 
Eqs.~(\ref{eq:F0}),(\ref{eq:F0_param}) and the Hartree approximation
\cite{s2d_hk}, and found that our result provides a much better matching.
Thus, solution (\ref{eq:F0}),(\ref{eq:F0_param}) will be used in further
calculation. In particular, we derive 
an approximate expression for the
derivative $dF/dr$, which will be useful in further analysis:
\begin{equation} \label{eq:Fderiv}
 {\left( \frac{d F}{d r} \right)}^2 \simeq 
         4 b_0^2 F^2 {\left( 1 - \frac{F}{F_m} \right)} .
\end{equation}

Now, after learning some important properties of the characteristic SH profile,
the next step is to determine parameters in the trial functions 
(\ref{eq:SH2_approx},\ref{eq:FF2_approx}). 
First, we substitute these expressions  
into the equation for the FF component in system (\ref{eq:uw2}). 
The resulting equality can't be satisfied exactly for any $r$, 
but we can use Eq.~(\ref{eq:Fderiv}) and get approximate relations 
between the parameters:
\begin{equation} \label{eq:FFeq}
 \begin{array}{l}
   {\displaystyle
    p b_s^2 \left[ 1 + 4 b_0^2 \left( p - 1 \right) \right] = 1, 
   } \\*[9pt] {\displaystyle
    p b_s^2 \left[ 1 + 4 b_0^2 \left( p - 1 \right) F_m^{-1} \right] = w_s, 
   } \end{array}
\end{equation}

Next, we fulfill the SH equation from (\ref{eq:uw2}) at $r \rightarrow 0$,
and obtain:
\begin{equation} \label{eq:SH2_eq0}
  w_s b_s^2 {\left( 1 - F_m \right)} - \alpha w_s 
       + \frac{1}{2} u_s^2 F_m^{2 p - 1} = 0.
\end{equation}
To determine the solution parameters, one more relation is needed,
preferably following from a conservation law, as this allows us to better
describe far-field asymptotics.
The difficulty here is that 
the equivalent dynamical system described by Eqs.~(\ref{eq:uw2}) is
dissipative, and thus is not Hamiltonian.
However, it is still possible to derive a condition similar
to Eq.~(\ref{eq:U_eq0}) for a planar configuration. 
To calculate an integral relation, 
we substitute approximate profiles 
(\ref{eq:SH2_approx},\ref{eq:FF2_approx}) into the equation for the SH
component in (\ref{eq:uw2}), 
multiply the equality by $d F(b_s r) / d r$,
and integrate over $(0,+\infty)$.
The resulting expression contains an averaged dissipation term, 
which can be found 
using Eq.~(\ref{eq:Feq})
\[
  \int_0^{+\infty} {\left( \frac{d F}{d r} \right)}^2 \frac{d r}{r} =
  \frac{1}{3} F_m^3 - \frac{1}{2} F_m^2, \nonumber
\]
and then the final equality is
\begin{equation} \label{eq:U2_eq0}
  w_s b_s^2 {\left( \frac{1}{2} - \frac{1}{3} F_m \right)} 
    - \frac{1}{2} \alpha w_s
    + \frac{1}{4 p + 2} u_s^2 F_m^{2 p - 1} = 0.
\end{equation}

Now all the parameters can be determined from the derived relations.
First, $p$ is found as a solution of cubic equation:
\begin{equation} \label{eq:2dp_p}
 \begin{array}{l}
   {\displaystyle
     8 \alpha b_0^2 p^3 
     + 2 \alpha \left( 1 - 6 b_0^2 \right) p^2
     + \left[ \left(4/3\right) F_m - 2 
    \right. } \\*[9pt] {\displaystyle \,\, \left.
              + \alpha \left( 4 b_0^2 - 1 \right) \right] p
     + \left[ 1 - \left(4/3\right) F_m \right] = 0 ,
   } \end{array}
\end{equation}
and then it is straightforward to calculate $b_s$, $w_s$, and $u_s$
one after another, employing Eqs.~(\ref{eq:FFeq},\ref{eq:SH2_eq0}).
Finally, using approximate expressions (\ref{eq:F0},\ref{eq:F0_param}) 
for the wave profiles at $\alpha=1$ and scaling transformations
Eqs.~(\ref{eq:SH2_approx},\ref{eq:FF2_approx}),
the solution can be written as:
\begin{equation} \label{eq:slv2d2}
 \begin{array}{l}
   {\displaystyle
    u_0(x) = u_m {\rm sech}^{2 p}(x b), 
   } \\*[9pt] {\displaystyle 
    w_0(x) = w_m {\rm sech}^{2}(x b),
   } \end{array}
\end{equation}
where $u_m = u_s F_m^p \sqrt{2}$, $w_m = w_s F_m$, and $b = b_s b_0$.
The amplitudes, characteristic widths, and scaling parameter $p$
are monotonic functions of $\alpha$, 
and for $1 \leq \alpha < +\infty$ change in the limits:
$1 \geq p > 1/2$, 
$b_0 \leq b < \sqrt{2 b_0^2 / (1-2 b_0^2)} \approx 1.448$, 
$(F_m \sqrt{2} \approx 3.363) \leq u_m < +\infty$,
$F_m \leq w_m < (F_m-2 b_0^2)/(1-2 b_0^2) \approx 5.267$.
Keep in mind however, that, 
unlike the (1+1)D case, the asymptotics for large
$\alpha$ are not exact. 

It is interesting to note that solution (\ref{eq:slv2d2}), 
describing the soliton profiles in bulk media, 
reduces to expressions (\ref{eq:slv2}) 
for the (1+1)D case if the parameters characterizing the
wave envelopes at $\alpha=1$ are chosen according to
Eq.~(\ref{eq:slv_al_1}) as $F_m = 3/2$ and $b_0 = 1/2$.

\begin{figure}
 \setlength{\epsfxsize}{8cm}
 \centerline{\mbox{\epsffile{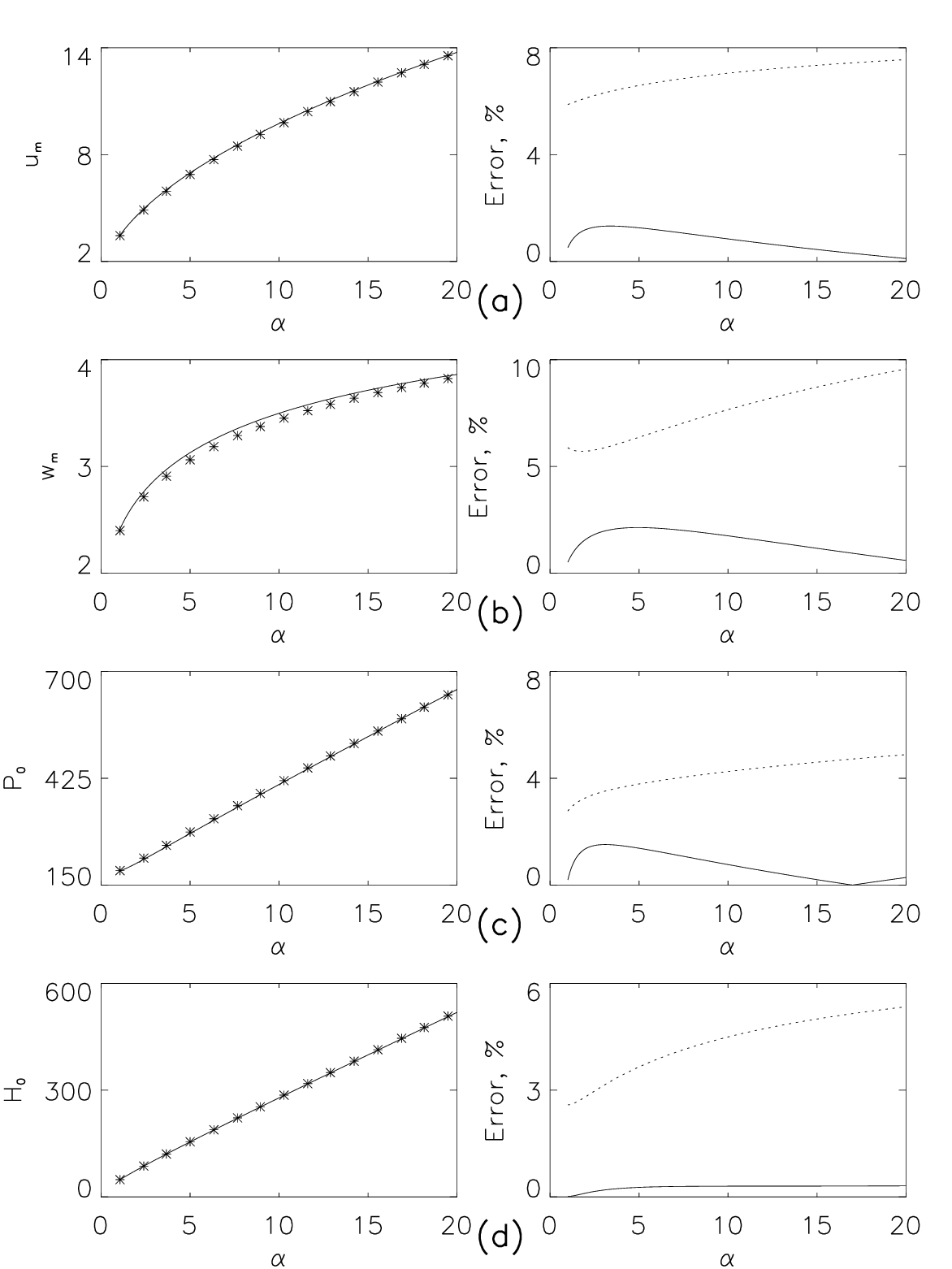}}}
 \vspace{2mm}
 \caption{ \label{fig:compar_2d2w}
   Comparison between exact numerical and approximate solutions,
   (\ref{eq:slv2d2}) and SGP, for solitons in bulk media. 
   The characteristics shown are the same as in Fig.~\ref{fig:compar_1d2w}.
 }
\end{figure}

\begin{figure}
 \setlength{\epsfxsize}{8.0cm}
 \centerline{\mbox{\epsffile{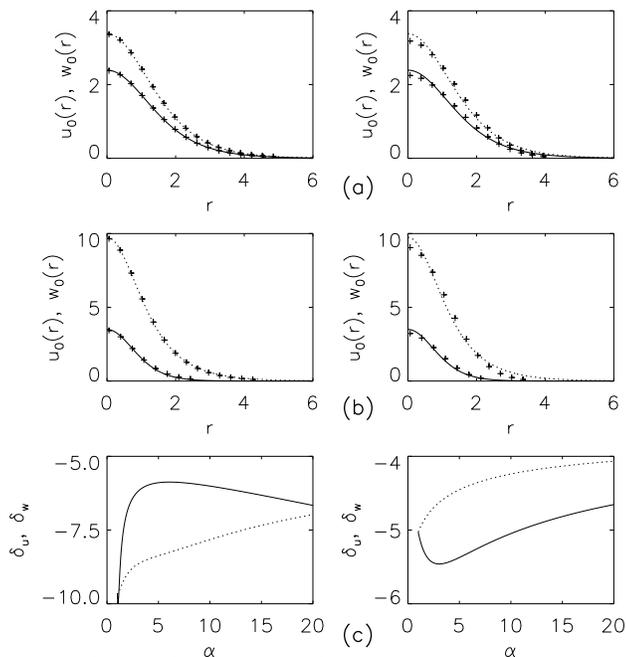}}}
 \vspace{2mm}
 \caption{ \label{fig:prof2d2w}
 Comparison of numerical results with analytical predictions for (2+1)D
 soliton profiles given by: 
 Eq.~(\ref{eq:slv2d2}) on the left, and the SGP on the right. 
 (a,b) Envelopes at $\alpha=1$ and $\alpha = 10$
 correspondingly. 
 (c)~Deviations of approximate profiles as defined in
     Eq.~(\ref{eq:uw2_err}).
 Notation is the same as in Fig.~\ref{fig:prof1d2w}.
}
\end{figure}

To study the accuracy of our solution, we made comparisons with direct
numerical calculations. The results are summarized in
Fig.~\ref{fig:compar_2d2w} in the same way as for solitons in planar
structures (see Fig.~\ref{fig:compar_1d2w}). 
We also look at relative deviations, defining them in a way similar to
(\ref{eq:uw_err}) as:
\begin{equation} \label{eq:uw2_err}
 \begin{array}{l}
  {\displaystyle 
   \delta_u = \ln { 
       \frac{\int_{-\infty}^{+\infty} 
             {\left| u_0^{\rm exact} - u_0^{\rm approx.} \right|}^2 r dr}
            {\int_{-\infty}^{+\infty} 
             {\left| u_0^{\rm exact} \right|}^2 r dr} 
       }, }
          \\*[15pt]
  {\displaystyle
   \delta_w = \ln {
         \frac{\int_{-\infty}^{+\infty} 
               {\left| w_0^{\rm exact} - w_0^{\rm approx.} \right|}^2 r dr} 
              {\int_{-\infty}^{+\infty} 
               {\left| w_0^{\rm exact} \right|}^2 r dr} 
       }.}
  \end{array}
\end{equation}
Each corresponding dependence is shown in Fig.~\ref{fig:prof2d2w}(c)
for our solution (\ref{eq:slv2d2}) on the left, and the SGP on the
right.
From the data presented, it follows that analytical solution
(\ref{eq:slv2d2}) provides a very good approximation for both integral
characteristics and soliton profiles [see also 
plots on the left in Figs.~\ref{fig:prof2d2w}(a) and \ref{fig:prof2d2w}(b)].
Especially accurate results are observed for mismatches of the order
$\alpha$ of $10^0$ to $10^1$. This range actually corresponds to interesting 
cases from an experimental point of view, as usually solitons are
observed more or less close to phase matching, i.e. 
$\alpha \sim 2 \sigma \simeq 4$ \cite{rev_stegeman,rev_chi2}.
The amazing precision is due to the 
fact that the approximate profile (\ref{eq:F0},\ref{eq:F0_param}) provides 
outstandingly good matching with the exact envelope, as shown in
Fig.~\ref{fig:prof2d2w}(a). Actually, it can be
claimed that this is an
{\em 'almost exact' solution at $\alpha=1$ for solitons in a bulk medium}, as
the deviations $\delta_u$ and $\delta_w$ become extremely small [see
the left plot in Fig.~\ref{fig:prof2d2w}(c)].
Note also that the approximation coefficients in \cite{s2d_hk} 
are not nearly as accurate as those given in (\ref{eq:F0_param}).

For larger mismatches, i.e. $\alpha > 10^1$, the
deviations increase, but the errors always remain smaller than for the SGP.
If even more accurate results are needed, it should be possible to start
the derivation with the NLSE in the limit $\alpha \rightarrow +\infty$, 
and then take into account terms of order $\alpha^{-1}$. 
As for the opposite case, $\alpha < 1$, linear tails may be expected to
form, similarly to a (1+1)D configuration. However, 
such special analysis is 
beyond the scope of current paper, 
and we believe that these open problems will be addressed in further studies.

\section{Concluding remarks}

We have studied the properties of two- and three-component quadratic 
bright solitons in planar waveguides and in bulk media.
Very accurate, yet simple and compact, approximate solutions 
have been derived to describe the actual wave profiles, with a precision
which the previously-known approximations could not achieve. 
Such amazingly good results have been obtained because the trial functions 
were chosen to correspond to the scaling properties of the solitons.
With the help of a specially-developed approach, 
the optimal values of fitting parameters have been found 
in a simple explicit form,
which wouldn't be possible with the variational method.

In particular, an {\em 'almost exact'} solution has been derived which 
{\em describes a whole family of two-wave planar solitons},
accounting for all the key properties. It not only provides perfect
estimations for integral characteristics (power and Hamiltonian), but
also closely matches the envelope profiles of both components 
for any mismatch.
On the other hand, the solution allowed us to 
reveal the existence of
non-oscillating {\em linear tails}, and rigorously describe their features,
which for example made it possible to explain and predict some
peculiarities of multi-soliton bound states.

We also considered three-wave coupling in an anisotropic medium between the
orthogonally-polarized FF and SH components. Mismatch values, when the
three wave solitons can exist, have been determined analytically,
and bifurcation scenarios have been described. Additionally,
an approximate solution for soliton profiles has been obtained, 
and it provides close estimations in a wide parameter region.

Quite interesting results have been obtained for the case of solitons in 
a bulk medium. Although exact analytical expressions for the two-wave 
profiles in a (2+1) dimensional case are not known, even in simpler limiting
cases (e.g., in the cascading limit the resulting single component NLSE 
is not integrable),
we have been able
to derive an 'almost exact' solution for a specific mismatch value.
General approximate expressions are also presented, giving very close
matching in a wide range of detunings, covering values close to phase
matching, which are of major interest from the experimental perspective.

\section*{Acknowledgements}

The author is grateful to Yu.S. Kivshar for initiating this project, 
and for fruitful discussions, and useful comments, 
and to A. Ankiewicz for a critical reading of this manuscript.

\end{multicols}

\begin{references}

\bibitem{rev_stegeman} 
For a comprehensive review, see 
G. Stegeman, D.J. Hagan, and L. Torner,  
Opt. Quantum Electron. {\bf 28}, 1691 (1996).

\bibitem{rev_altern}
M. Georgieva-Grosse and A. Shivarova, p. 319; 
V.M. Agranovich and A.M. Kamchatnov, p.277, 
in {\em Advanced Photonics with Second-Order Optically 
Nonlinear Processes},  edited by A.D. Boardman, L. Pavlov, and S. Tanev, Eds. 
(Kluwer, Dordretch, 1998)

\bibitem{BEC}
P.D. Drummond, K.V. Kheruntsyan, and H. He,
Phys. Rev. Lett. {\bf 81}, 3055 (1998).

\bibitem{rev_chi2} 
For an overview of quadratic spatial solitons, 
see L. Torner, in {\em Beam Shaping and Control with Nonlinear Optics}, 
F. Kajzer and R. Reinisch, Eds. (Plenum, New York, 1998), p. 229;  
Yu.S. Kivshar, in {\em Advanced Photonics with Second-Order Optically 
Nonlinear Processes},  A.D. Boardman, L. Pavlov, and S. Tanev, Eds. 
(Kluwer, Dordretch, 1998), p. 451.

\bibitem{var_m}
D. Anderson, Phys. Rev. A {\bf 27}, 3135 (1983).

\bibitem{aa}
A. Ankiewicz, N. Akhmediev, G.D. Peng, and P.L. Chu,
Opt. Comm. {\bf 103}, 410 (1993).

\bibitem{f_ks} 
Yu.N. Karamzin and A.P. Sukhorukov,
Pis'ma Zh. Eksp. Teor. Fiz. {\bf 20}, 734 (1974)
 [JETP Lett. {\bf 20}, 339 (1974)];
Zh. Eksp. Teor. Fiz. {\bf 68}, 834 (1975)
 [Sov. Phys. JETP {\bf 41}, 414 (1976)].

\bibitem{f_ksf}
Yu.N. Karamzin, A.P. Sukhorukov, and T.S. Filipchuk,
 Vest. Mosk. Univ. {\bf 33}, 73 (1978)
 [Moscow Univ. Phys. Bull. {\bf 19}, 91 (1978)].

\bibitem{sfam_bk}
A.V. Buryak and Yu.S. Kivshar, 
Opt. Lett. {\bf 19}, 1612 (1994); {\bf 20}, 1080(E) (1995);
Phys. Lett. A {\bf 197}, 407 (1995).

\bibitem{krub}
A.A. Kanashov and A.M. Rubenchik, Physica D {\bf 4}, 122 (1981).

\bibitem{sfam_torner}
L. Torner, C.R. Menyuk, and G. Stegeman, Opt. Lett. {\bf 19}, 1615 (1994);
L. Torner, Opt. Comm. {\bf 114}, 136 (1995).


\bibitem{sfam_drum} 
H. He, M.J. Werner, and P.D. Drummond, Phys. Rev. E {\bf 54}, 896 (1996).

\bibitem{stability} 
D.E. Pelinovsky, A.V. Buryak, and Yu.S. Kivshar, 
Phys. Rev. Lett. {\bf 75}, 591 (1995).

\bibitem{nls_drum}
M.J. Werner and P.D. Drummond, Opt. Lett. {\bf 19}, 613 (1994).

\bibitem{nls_berge}
L. Berg\'e, V.K. Mezentsev, J.J. Rasmussen, and J. Wyller,
Phys. Rev. A {\bf 52}, R28 (1995).

\bibitem{var_exp} 
V.V. Steblina, Yu.S. Kivshar, M. Lisak, and B.A. Malomed, 
Opt. Commun. {\bf 118}, 345 (1995).

\bibitem{var_exp_bs} 
A.D. Boardman, K. Xie, and A. Sangarpaul, 
Phys. Rev. A {\bf 52}, 4099 (1995).

\bibitem{var_sech} 
V.M. Agranovich {\em et al.}, Phys. Rev. B {\bf 53}, 15451 (1996);
V.M. Agranovich {\em et al.}, Phys. Rev. E {\bf 55}, 1894 (1997);
C. Balslev Clausen, P.L. Christiansen, and L. Torner, 
Opt. Comm. {\bf 136}, 185 (1997).

\bibitem{bs_bur}
A.V. Buryak, Phys. Rev. E {\bf 52}, 1156 (1995).

\bibitem{bs_ml}
D. Mihalache, F. Lederer, D. Mazilu, and \mbox{L.-C.} Crasovan,
Opt. Eng. {\bf 35}, 1616 (1996).

\bibitem{bs_champ}
A.C. Yew, A.R. Champneys, and P.J. McKenna,
J. Nonlinear Sci. {\bf 9}, 33 (1999).

\bibitem{us} 
Yu.S. Kivshar, A.A. Sukhorukov, and S.M. Saltiel, 
{\em Two-color multistep cascading and parametric soliton-induced
waveguides}, to be published; e-print \mbox{patt-sol/9905002}.

\bibitem{s2d_hk}
K. Hayata and M. Koshiba, Phys. Rev. Lett. {\bf 71}, 3275 (1993);
{\bf 72}, 178(E) (1994).

\bibitem{s2d_fam}
A.V. Buryak, Yu.S. Kivshar, and V.V. Steblina, 
Phys. Rev. A {\bf 52}, 1670 (1995).


\end{references}
\end{document}